\documentclass{article}%

\usepackage{amssymb}
\usepackage{makeidx}
\usepackage{amsfonts}
\usepackage{amsmath}
\usepackage{graphicx}

\begin{document}

\title{From Objective Amplitudes to Bayesian Probabilities\thanks{Invited paper
presented at the International Conference on Foundations of Probability and
Physics -- 4 (V\"{a}xj\"{o} University, Sweden, 2006).}}
\author{Ariel Caticha\\{\small Department of Physics, University at Albany-SUNY, }\\{\small Albany, NY 12222, USA (ariel@albany.edu). }}
\date{}
\maketitle

\begin{abstract}
We review the Consistent Amplitude approach to Quantum Theory and argue that
quantum probabilities are explicitly Bayesian. In this approach amplitudes are
tools for inference. They codify objective information about how complicated
experimental setups are put together from simpler ones. Thus, probabilities
may be partially subjective but the amplitudes are not.

\end{abstract}

\section{Introduction}

Whether Bayesian probabilities -- degrees of belief -- are (1) subjective or
(2) objective or (3) somewhere in between, is still a matter of controversy. I
vote for (3). Probabilities will always retain a \textquotedblleft
subjective\textquotedblright\ element because translating information into
probabilities involves judgments and different people will inevitably judge
differently. On the other hand, not all probability assignments are equally
useful and it is plausible that what makes some assignments better than others
is that they represent or reflect some objective feature of the world. One
might even say that what makes them better is that they are closer to the
\textquotedblleft truth\textquotedblright. Thus, probabilities can be
characterized by both subjective and objective elements and, ultimately, it is
their objectivity that makes probabilities useful.

Quantum mechanics adds an interesting twist because the recipe for calculating
quantum probabilities is inflexible: apply Born's rule to the amplitudes.
There seems to be no room for judgments here. One possibility is that the
objective nature of the wave function extends also to the quantum
probabilities. In this view quantum and Bayesian probabilities differ in an
essential way: quantum probabilities are not Bayesian. A second possibility is
that it is the other way around and it is the subjectivity of the
probabilities that infects the wave functions. The purpose of this paper is to
argue for a third possibility.

Our goal here is to review the main ideas behind the Consistent Amplitude
approach to Quantum Theory (CAQT) \cite{Caticha98a}-\cite{Caticha98c} in order
to argue that there is no need to distinguish between quantum and classical
probabilities. In the CAQT approach probabilities are explicitly Bayesian --
they reflect the degree to which we believe that detectors will successfully
detect -- and yet, there is nothing subjective about the wave function that
conveys the relevant information about the (idealized) experimental setup. The
situation here is quite analogous to assigning Bayesian probabilities to the
outcomes of a die toss based on the objective information that the (idealized)
die is a perfectly symmetric cube. The probabilities may be partially
subjective, but the information on which they are based is intended to
represent an objective feature of the world.

Many discussions on the foundations of quantum theory start from the abstract
mathematical formalism of Hilbert spaces and some postulates\ or rules
prescribing how the formalism should be used. Their goal is to discover a
suitable interpretation. The CAQT is different in that it proceeds in the
opposite direction: first one specifies the objective the theory is meant to
accomplish and then a suitable formalism is derived from a set of
\textquotedblleft reasonable\textquotedblright\ assumptions. In a sense, the
interpretation is the starting point and the formalism is derived from it.

The objective of the CAQT is to predict the outcomes of certain idealized
experiments on the basis of information about how complicated experimental
setups are put together from simpler ones. The theory is, by design, a theory
of inference from available information. The \textquotedblleft
reasonable\textquotedblright\ assumptions are four. The first specifies the
kind of setups about which we want to make predictions. The second assumption
establishes what is the relevant information on the basis of which the
predictions will be made and how this information is codified. It is at this
stage that amplitudes and wave functions are introduced as tools for the
consistent manipulation of information. The third and fourth assumptions
provide the link between the amplitudes and the actual prediction of
experimental outcomes. Although none of these assumptions refer to
probabilities, all the elements of quantum theory, including indeterminism and
the Born rule, Hilbert spaces, linear and unitary time evolution, are derived.
\cite{Caticha98a}-\cite{Caticha98c}

\section{Setups and amplitudes}

The first and most crucial step is a decision about the subject matter. We
choose a pragmatic, operational approach: the objective of quantum theory is
to predict the outcomes of experiments carried out with certain idealized setups.

Next we note that if setups are related in some way -- perhaps two setups can
be connected to build a third one -- then information about one may be
relevant to predictions about the other. This suggests that one should
identify the possible relations among setups. To avoid irrelevant technical
distractions we consider a very simple quantum system, a \textquotedblleft
particle\textquotedblright\ that lives on a discrete lattice and has no spin
or other internal structure. (In fact, whether the system is a particle or a
wave or both or neither is immaterial; the word `particle' is selected only
because a word, some word, must be used to refer to the system.) The
generalization to more complex configuration spaces should be straightforward.%

\begin{figure}[ptb]
\begin{center}
\includegraphics[width=0.6\textwidth]{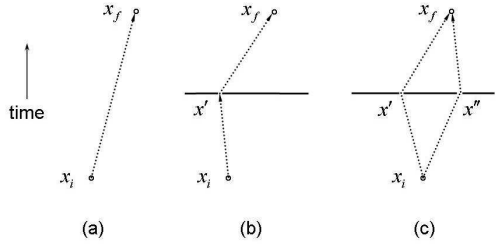}
\caption{(a) The simplest setup $[x_{f},x_{i}]$. (b) Two successive setups
$[x_{f},x^{\prime}]$ and $[x^{\prime},x_{i}]$ are combined with the
\textsc{and} operation. (c) Two setups in parallel, $[x_{f},x^{\prime},x_{i}]$
and $[x_{f},x^{\prime\prime},x_{i}]$ are combined with the \textsc{or}
operation.}%
\end{center}
\end{figure}

The simplest experimental setup, denoted by $[x_{f},x_{i}]$, is shown in Fig.
1(a). It consists of a source at the space-time point $x_{i}=(\vec{x}%
_{i},t_{i})$ and a detector at $x_{f}=(\vec{x}_{f},t_{f})$. Both $\vec{x}_{i}
$ and $\vec{x}_{f}$ are assumed to be points on a discrete lattice and the
intermediate region may be filled by various kinds of fields. In Fig. 1(b) we
show a slightly more complex setup that we denote by $[x_{f},x^{\prime}%
,x_{i}]$ where the source is at $x_{i}$, the detector at $x_{f}$, and we have
introduced a \textquotedblleft filter\textquotedblright\ that blocks all paths
from $x_{i}$ to $x_{f}$ except via the intermediate point $x^{\prime}$. The
possibility of introducing many filters each with several holes (as in Fig.
1c) leads to allowed setups of the general form $a=[x_{f},s_{n},s_{n-1}%
,\ldots,s_{2},s_{1},x_{i}]$ where each $s_{n}=(x_{n},x_{n}^{\prime}%
,x_{n}^{\prime\prime},\ldots)$ represents a filter at time $t_{n}$,
intermediate between $t_{i}$ and $t_{f}$, with multiple holes at $\vec{x}%
_{n},\vec{x}_{n}^{\prime},\vec{x}_{n}^{\prime\prime},\ldots$

There are two basic kinds of relations among setups. The first, called
\textsc{and}, arises when two setups $a$ and $b$ are placed in immediate
succession resulting in a third setup which we denote by $ab$. It is necessary
that the destination point of the earlier setup coincide with the source point
of the later one, otherwise the combined $ab$ is not allowed. For example,
$[x_{f},x^{\prime}][x^{\prime},x_{i}]=[x_{f},x^{\prime},x_{i}]$ is shown in
Fig. 1(b). With the single setup $[x_{f},x^{\prime},x_{i}]$ we can perform
several different experiments. We can place a source at $x_{i}$ and the
detector at $x^{\prime}$ in which case we are only using the $[x^{\prime
},x_{i}]$ component; or we can place the source at $x^{\prime}$ and the
detector at $x_{f}$ and use only the $[x_{f},x^{\prime}]$ component; or we can
place a source at $x_{i}$, the detector at $x_{f}$, and use the whole thing
$[x_{f},x^{\prime},x_{i}]$. Although we can only do one of these experiments
at a time, the three experiments are clearly related and knowing something
about one setup may be helpful in making predictions about another.

The second relation, called \textsc{or}, arises from the possibility of
opening additional holes in any given filter. Specifically, when (and
\emph{only} when) two setups $a^{\prime}$ and $a^{\prime\prime}$ are identical
except on one single filter where none of the holes of $a^{\prime}$ overlap
any of the holes of $a^{\prime\prime}$, then we may form a third setup $a$,
denoted by $a^{\prime}\vee a^{\prime\prime}$, which includes the holes of both
$a^{\prime}$ and $a^{\prime\prime}$. The example
\begin{align}
\lbrack x_{f},(x^{\prime},x^{\prime\prime}),x_{i}]  & =[x_{f},x^{\prime}%
,x_{i}]\vee\lbrack x_{f},x^{\prime\prime},x_{i}]\\
& =([x_{f},x^{\prime}][x^{\prime},x_{i}])\vee([x_{f},x^{\prime\prime
}][x^{\prime\prime},x_{i}])
\end{align}
is shown in Fig. 1(c). Again we see that we can perform a wide variety of
\emph{related} experiments with the same setup by judicious placement of
source and detector and by blocking or not the appropriate holes.

Provided the relevant setups are allowed the basic properties of \textsc{and}
and \textsc{or} are quite obvious: \textsc{or} is commutative, but
\textsc{and} is not; both \textsc{and} and \textsc{or} are associative, and
finally, \textsc{and} distributes over \textsc{or} but not vice-versa. One
should emphasize that these are physical rather than logical connectives. They
represent our idealized ability to construct more complex setups out of
simpler ones and they differ substantially from their Boolean and quantum
logic counterparts. In Boolean logic not only \textsc{and} distributes over
\textsc{or} but \textsc{or} also distributes over \textsc{and}, while in
quantum logic propositions refer to quantum properties at one time rather than
to processes in time.

Thus, our first assumption is

\begin{itemize}
\item[\textbf{A1.}] The goal of quantum theory is to predict the outcomes of
experiments involving setups built from components connected through
\textsc{and} and \textsc{or}.
\end{itemize}

Having identified relations among setups we now seek a way to handle them
quantitatively. A representation of \textsc{and/or} can be obtained by
assigning a complex number $\phi(a)$ to each setup $a$ in such a way that
relations among setups translate into relations among the corresponding
complex numbers. What gives the theory its robustness, its uniqueness, is the
requirement that the assignment be consistent: if there are two different ways
to compute $\phi(a)$ the two answers must agree. The remarkable consequence of
requiring consistency is embodied in the following

\noindent\textbf{Regraduation Theorem: }Given one consistent representation of
\textsc{and/or} in terms of complex numbers $\phi(a)$, one can
\textquotedblleft regraduate\textquotedblright\ with an appropriately chosen
function $\psi$ to obtain an equivalent and more convenient assignment,
$\psi(a)=\psi(\phi(a))$, so that \textsc{and} and \textsc{or} are respectively
represented by multiplication and addition,
\begin{equation}
\psi\left(  ab\right)  =\psi\left(  a\right)  \psi\left(  b\right)
\quad\text{and\quad}\psi\left(  a\vee a^{\prime}\right)  =\psi\left(
a\right)  +\psi\left(  a^{\prime}\right)  \text{.}\label{SP rules}%
\end{equation}
Complex numbers assigned in this way will be called \emph{amplitudes}. The
theorem is proved in \textbf{\cite{Caticha98a, Caticha98b}}. (For an
independent earlier derivation see ref. \cite{Tikochinsky88}.)

It would be interesting to explore the consequences of keeping track of more
information about the setups by using representations of \textsc{and/or} in
terms of more complicated mathematical objects. It is likely that the
resulting theory would differ from quantum mechanics in important ways. For
our current purposes however we restrict ourselves to a representation in
terms of amplitudes. This is our second assumption,

\begin{itemize}
\item[\textbf{A2.}] The relevant information for predicting the outcome of an
experiment with a setup $a$ is codified into its complex amplitude $\psi(a)$.
\end{itemize}

\section{Wave functions and time evolution}

The observation, already expressed by Feynman \cite{Feynman48}, that leads to
the fundamental evolution equation (and eventually to path integrals) is that
a filter that is totally covered with holes is equivalent to having no filter
at all,
\begin{equation}
\lbrack x_{f},x_{i}]=\bigvee_{\text{all}\,\,\vec{x}\,\text{at}\,t}%
([x_{f},x_{t}][x_{t},x_{i}])~.
\end{equation}
In terms of the corresponding amplitudes, this is quantitatively expressed as
\begin{equation}
\psi(x_{f},x_{i})=\sum_{\text{all}\,\vec{x}\,\text{at}\,t}\psi(x_{f}%
,x_{t})\,\psi(x_{t},x_{i}).
\end{equation}

Since we are interested only in the response of the detectors we find that in
there are situations where amplitudes keep track of too much information.
Indeed, note that there are many possible combinations of starting points
$x_{i}$ and of interactions prior to the time $t$ that will result in
identical evolution after time $t$. What these different possibilities have in
common is that they all lead to \hspace{0pt}the same numerical value for the
amplitude $\psi(x_{t},x_{i})$. It is therefore convenient to set $\Psi(\vec
{x},t)=\psi(x_{t},x_{i})$ and omit all reference to the irrelevant starting
point $x_{i}$. The object thus introduced, $\Psi(\vec{x},t)$, is called a wave
function. The usual language is that $\Psi(\vec{x},t)$ describes the state of
the particle at time $t$. In the CAQT approach we say that $\Psi(\vec{x},t)$
encodes information about those features of the setup prior to $t$ that are
relevant to time evolution after $t$.

The evolution equation can then be written as%

\begin{equation}
\Psi(\vec{x}_{f},t_{f})=\sum_{\text{all }\,\vec{x}\,\text{at}\,t}\psi(\vec
{x}_{f},t_{f};\vec{x},t)\,\Psi(\vec{x},t)\text{,}\label{evolution}%
\end{equation}
\newline which is equivalent to a linear Schr\"{o}dinger equation as can
easily be seen \cite{Caticha98a}\cite{Caticha98b} by differentiating with
respect to $t_{f}$ and evaluating at $t_{f}=t$. Thus, a quantum theory
formulated in terms of consistently assigned amplitudes must be linear:
nonlinear modifications of quantum mechanics must either violate assumptions
\textbf{A1} or \textbf{A2} or else be internally inconsistent.

\section{Hilbert space}

In the usual approach to quantum theory the linearity embodied in the
principle of superposition and the linearity of the Schrodinger equation seem
completely unrelated. Indeed, all proposals for a non-linear quantum mechanics
have invariably maintained the former and broken the latter. Within the CAQT
approach this is not allowed. One cannot preserve the linearity of
superpositions and break the linearity of evolution because both follow from
the same consistency requirement that led to the sum and product rules for amplitudes.

The superposition principle states that if a system can be prepared in a state
$\Psi^{\prime}$ and it can be prepared in another state $\Psi^{\prime\prime}$
then it can also be prepared in a linear superposition $\Psi=\alpha
\Psi^{\prime}+\beta\Psi^{\prime\prime}$. An interesting question is whether
the superposition can actually be prepared. Given a generic pair of states
$\Psi^{\prime}$ and $\Psi^{\prime\prime}$, what is the specific setup that
will prepare the state $\alpha\Psi^{\prime}+\beta\Psi^{\prime\prime}$?

Within the CAQT approach this annoying question does not arise because we are
not interested in preparing arbitrary states, but only in predicting the
outcomes of experiments with those special setups described in assumption
\textbf{A1}. For these special setups it is easy to see that linear
superpositions do arise. For example, let $\Psi^{\prime}(\vec{x},t)=\psi
(\vec{x},t;\vec{x}^{\prime},t_{0})$ be the wave function at time $t$ when the
source is at $(\vec{x}^{\prime},t_{0})$, and let $\Psi^{\prime\prime}(\vec
{x},t)=\psi(\vec{x},t;\vec{x}^{\prime\prime},t_{0})$ be the wave function when
the source is at $(\vec{x}^{\prime\prime},t_{0})$. See Fig. 2. Linear
superpositions of $\Psi^{\prime}(\vec{x},t)$ and $\Psi^{\prime\prime}(\vec
{x},t)$ can be prepared by placing a filter at $t_{0}$ with holes at $\vec
{x}^{\prime}$ and $\vec{x}^{\prime\prime}$ and a source at a point $(\vec
{x}_{i},t_{i})$ prior to $t_{0}$. Then the overall amplitude for this new
setup, $\psi(\vec{x},t;\vec{x}_{i},t_{i})$, is
\begin{equation}
\Psi(\vec{x},t)=\alpha\Psi^{\prime}(\vec{x},t)+\beta\Psi^{\prime\prime}%
(\vec{x},t)\text{ ,}%
\end{equation}
where $\alpha=\psi(\vec{x}^{\prime},t_{0};\vec{x}_{i},t_{i})$ and $\beta
=\psi(\vec{x}^{\prime\prime},t_{0};\vec{x}_{i},t_{i})$. Notice that the
complex numbers $\alpha$ and $\beta$ can be changed at will by moving the
starting point $(\vec{x}_{i},t_{i})$ or by introducing additional fields
between $t_{i}$ and $t_{0}$.%

\begin{figure}[ptb]
\begin{center}
\includegraphics[width=.55\textwidth]{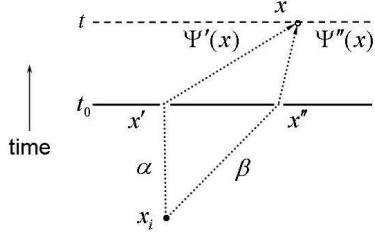}
\caption{Illustrating how the superposition principle follows from the sum and
product rules for consistent amplitudes.}%
\end{center}
\end{figure}

Thus, as a consequence of the sum and product rules, eq.(\ref{SP rules}),
linear superpositions of wave functions arise naturally and we can ask what
else is needed for them to form a Hilbert space. The basic question is whether
an equally compelling inner product is available. The answer is quite
illuminating because of the support it gives to viewing quantum mechanics as a
theory of inference rather than as a law of nature.

An inner product defines lengths and angles. We find that without any
additional assumptions the basic components of setups, the filters, take us a
long way toward an inner product because they already supply us with a concept
of orthogonality. The action of a filter $P$ with holes at a set of points
$\vec{x}_{p}$ is to turn the wave function $\Psi(\vec{x})$ into the wave
function $P\Psi(\vec{x})=\sum_{p}\delta_{\vec{x},\vec{x}_{p}}\Psi(\vec{x})$.
Since the filters $P$ act as projectors, $P^{2}=P$, it follows that any given
filter defines two special classes of wave functions. One is the subspace of
wave functions that are transmitted by the filter without any changes, and the
other is the subspace of wave functions that are completely blocked. We will
\emph{define} these two subspaces as being orthogonal to each other.

Any wave function can be decomposed into orthogonal components, $\Psi=\Psi
_{P}+\Psi_{1-P}$ where $\Psi_{P}=P\Psi$ is unaffected by the filter and
$\Psi_{1-P}=(1-P)\Psi$ is blocked: $P\Psi_{P}=\Psi_{P}$ and $P\Psi_{1-P}=0$. A
particularly convenient expansion in orthogonal components is that defined by
a complete set of elementary filters. A filter $P_{i}$ is elementary if it has
a single hole at $\vec{x}_{i}$. It acts by multiplying $\Psi(\vec{x})$ by
$\delta_{\vec{x},\vec{x}_{i}}$. The set is complete if $\sum_{i}P_{i}=1$. Then
$\Psi(\vec{x})=\sum_{i}A_{i}\,\delta_{\vec{x},\vec{x}_{i}}$, where $A_{i}%
=\Psi(\vec{x}_{i})$.

At this point it is convenient to introduce the familiar Dirac notation.
Instead of $\Psi(\vec{x})$ and $\delta_{\vec{x},\vec{x}_{i}}$ we shall write
$|\Psi\rangle$ and $|i\rangle$, so that $|\Psi\rangle=\sum_{i}A_{i}|i\rangle$.
The question is what else, in addition to orthogonality, do we need to define
the inner product. Recall that an inner product satisfies three conditions:
(a) $\langle\Psi|\Psi\rangle\geqslant0$ with $\langle\Psi|\Psi\rangle=0$ if
and only if $|\Psi\rangle=0$, (b) linearity in the second factor, and (c)
antilinearity in the first factor, $\langle\Phi|\Psi\rangle=\langle\Psi
|\Phi\rangle^{\ast}$. Conditions (b) and (c) define the product of
$|\Phi\rangle=\sum_{j}B_{j}|j\rangle$ with $|\Psi\rangle=\sum_{i}%
A_{i}|i\rangle$ in terms of the product of $|j\rangle$ with $|i\rangle$,
\begin{equation}
\langle\Phi|\Psi\rangle=\sum_{i}B_{j}^{\ast}A_{i}\langle j|i\rangle~.
\end{equation}
The orthogonality of the basis functions $\delta_{\vec{x},\vec{x}_{i}}$ is
naturally encoded into the inner product by setting $\langle j|i\rangle=0$ for
$i\neq j$, but the case $i=j$ remains undetermined, constrained by condition
(a) to be real and positive. Clearly an additional assumption is necessary.
Here it is:

\begin{itemize}
\item[\textbf{A3.}] If there is no reason to prefer one region of
configuration space over another they should be assigned equal a priori weight.
\end{itemize}

\noindent We therefore choose $\langle i|i\rangle$ equal to a constant which,
without losing generality, we set equal to one:
\begin{equation}
\langle i|j\rangle=\delta_{ij}\Rightarrow\langle\Phi|\Psi\rangle=\sum_{i}%
B_{i}^{\ast}A_{i},
\end{equation}
and this leads to the Hilbert norm $\left\Vert \Psi\right\Vert ^{2}%
=\langle\Psi|\Psi\rangle=\sum_{i}|A_{i}|^{2}$. The significance of \textbf{A3}
will be better understood once we explore its implications in the next section.

\section{Probabilities: the Born rule}

Finally, we get to the question of using the information encoded into
amplitudes to predict the outcomes of experiments. Our argument invokes time
evolution, eq. (\ref{evolution}), in an essential way. This is in contrast to
other derivations of the Born rule (such as Gleason's) which are essentially static.

It is perhaps best to start with a simple special case. Suppose that the
preparation procedure is such that $\Psi(\vec{x},t)$ vanishes at a certain
point $(\vec{x}_{0},t)$. Then, according to eq. (\ref{evolution}), placing an
obstacle at the single point $(\vec{x}_{0},t)$ (\textit{i.e.}, placing a
filter at $t$ with holes everywhere except at $\vec{x}_{0}$) has no effect on
the subsequent evolution of $\Psi$. \emph{Since the mathematical relations
among amplitudes are meant to reflect the corresponding physical relations
among setups, it seems natural to assume that the presence or absence of the
obstacle at }$\vec{x}_{0}$\emph{\ will have no observable effects.} What if
the obstacle was itself a detector? It is just as natural to assume that since
the obstacle/detector inflicts no action on the system, then it must not
itself suffer any reaction. Therefore, we conclude that when $\Psi(\vec{x}%
_{0},t)=0$ the particle will not be detected at $(\vec{x}_{0},t)$.

Convincing and natural as this argument might be, it is important to recognize
that an assumption was made: no amount of logic could conceivably bridge the
gap from a mathematical representation in terms of amplitudes to the
prediction of an actual physical event. The assumption can be generalized to
the following general interpretative rule:

\begin{itemize}
\item[\textbf{A4.}] If the introduction at time $t$ of a filter blocking those
components of the wave function characterized by a certain property
$\mathcal{P}$ has no effect on the future evolution of a particular wave
function $\Psi(t)$ then when the wave function happens to be $\Psi\left(
t\right)  $ the property $\mathcal{P}$ will not be detected.
\end{itemize}

\noindent Note that \textbf{A4} deals with a situation of complete certainty;
no probabilities are mentioned.

The deduction of the Born rule now proceeds as in ref. \cite{Caticha98b}.
Briefly the idea is as follows. We want to predict the outcome of an
experiment in which a detector is placed at a certain $\vec{x}_{k}$ when the
system is in state $|\Psi\rangle=\sum_{i}A_{i}|i\rangle$. In \cite{Caticha98b}
we showed that the state for an ensemble of $N$ identically prepared,
independent replicas of our particle is the product $|\Psi_{N}\rangle
=\prod_{\alpha=1}^{N}|\Psi_{\alpha}\rangle$. Now we apply the interpretative
rule \textbf{A4}. Suppose that in the $N$-particle configuration space we
place a special filter, denoted by $P_{f,\varepsilon}^{k}$, the action of
which is to block all components of $|\Psi_{N}\rangle$ except those for which
the fraction $n/N$ of replicas at $\vec{x}_{k}$ lies in the range from
$f-\varepsilon$ to $f+\varepsilon$. The difference between the states
$P_{f,\varepsilon}^{k}|\Psi_{N}\rangle$ and $|\Psi_{N}\rangle$ is measured by
the relative Hilbert distance, \linebreak$||P_{f,\varepsilon}^{k}|\Psi
_{N}\rangle-|\Psi_{N}\rangle||^{2}/\langle\Psi_{N}|\Psi_{N}\rangle$. The
result of this calculation is \cite{Caticha98b} \
\begin{equation}
\underset{N\rightarrow\infty}{\lim}\,\left\Vert P_{f,\varepsilon}^{k}|\Psi
_{N}\rangle-|\Psi_{N}\rangle\right\Vert ^{2}=1-\int_{f-\varepsilon
}^{f+\varepsilon}\delta\left(  f^{\prime}-|A_{k}|^{2}\right)  df^{\prime
}\text{ ,}\label{difference}%
\end{equation}
where we have normalized $\langle\Psi|\Psi\rangle=\langle\Psi_{N}|\Psi
_{N}\rangle=1$.

We see that for large $N$ the filter $P_{f,\varepsilon}^{k}$ has no effect on
the state $|\Psi_{N}\rangle$ provided $f$ lies in a range $2\varepsilon$ about
$|A_{k}|^{2}$. Therefore, according to \textbf{A4}, the state $|\Psi
_{N}\rangle$ does not contain any fractions outside this range. On choosing
stricter filters with $\varepsilon\rightarrow0$ we conclude that we can
predict with complete certainty that detection at $\vec{x}_{k}$ will occur for
a fraction $|A_{k}|^{2}$ of the replicas and that detection will not occur for
the remaining fraction $1-|A_{k}|^{2}$. For any one of the \emph{identical}
individual replicas there is no such certainty; the best one can do is to say
that detection will occur with a certain probability $\Pr(k)$. In order to be
consistent with the law of large numbers the assigned value must agree with
the Born rule
\begin{equation}
\Pr(k)=|A_{k}|^{2}\,.\label{Born rule}%
\end{equation}

The important role played by the inner product, introduced through assumption
\textbf{A3}, into eq.(\ref{difference}) must be emphasized. In fact, had we
weighted the $|i\rangle$'s differently and chosen a different normalization,
say $\langle i|i\rangle=w_{i}$, the resulting probability would have been
$\Pr(i)$ $=w_{i}|A_{i}|^{2}$ instead of eq.(\ref{Born rule}).

It is instructive to explore this issue further. Nothing in our choice of a
discrete lattice requires it to be uniform and periodic. Indeed, a non-uniform
discrete lattice could have been defined in terms of an arbitrary frame of
curvilinear coordinates. If the underlying configuration space is Euclidean
the use of curvilinear rather than cartesian coordinates is a matter of choice
but in the general case of curved spaces curvilinear coordinates are unavoidable.

Assumption \textbf{A3} suggests\textbf{\ }that each cell $i$ in the
non-uniform lattice be weighted by its own volume which we will denote by
$w_{i}=g_{i}^{1/2}\Delta x$. (The usual notation for a volume element in
arbitrary coordinates is $g^{1/2}(x)dx$ where $g(x)$ is the determinant of the
metric tensor, and $dx=dx^{1}\ldots dx^{3}$ in three dimensions.) In the
continuum limit \cite{Caticha98c} we let $\Delta x\rightarrow dx$. Replacing
$w_{i}^{-1}|i\rangle=(g_{i}^{1/2}\Delta x)^{-1}|i\rangle$ by$\,|\vec{x}%
\rangle$ the completeness condition $1=\sum_{i}P_{i}=\sum_{i}w_{i}%
^{-1}|i\rangle\langle i|$ becomes $1=\int g^{1/2}dx\,|\vec{x}\rangle
\langle\vec{x}|$. Next, replace $\delta_{ij}/\Delta x$ by $\delta(\vec{x}%
-\vec{x}^{\prime})$ and the inner product $\langle i|j\rangle=w_{i}\delta
_{ij}$ becomes $\langle\vec{x}|\vec{x}^{\prime}\rangle=g^{-1/2}\delta(\vec
{x}-\vec{x}^{\prime})$. Finally, replace $A_{i}$ by $A(\vec{x})$ and the state
$|\Psi\rangle=\sum_{i}A_{i}|i\rangle$ becomes $|\Psi\rangle=\int
g^{1/2}dx\,A(\vec{x})\,|\vec{x}\rangle$. The Born rule, eq. (\ref{Born rule}),
becomes
\begin{equation}
\Pr(dx)=g^{1/2}dx\,|A(\vec{x})|^{2}%
\end{equation}
which shows that $|A(\vec{x})|^{2}$ is the probability \emph{density}.

It is sometimes argued that while there is an element of subjectivity in the
nature of classical probabilities but that quantum probabilities are
different, that they are totally objective because they are given by $|A|^{2}
$. We have just shown that this assignment is neither more nor less subjective
than say, assigning probabilities to each face of a die. Just like we assign
probability $1/6$ when we have no reason to prefer one face of the die over
another, the Born rule follows, even in curved spaces, when we have no reason
to prefer one volume element over another --- provided their volumes are equal.

\section{Conclusion}

Probabilities may be subjective but the information on the basis of which we
update probabilities is not supposed to be. Indeed, we want to update from a
prior to a posterior distribution because we believe the posterior
distribution will give us a better assessment of truth. Posterior
distributions are \textquotedblleft objectively\textquotedblright\ better than
prior distributions.

We have argued that amplitudes and wave functions codify objective information
about the way experimental setups are put together. Quantum probabilities take
this information into account and thereby incorporate an objective element --
this is why they work. On the other hand, although amplitudes and wave
functions may be representations of objective information, an element of
subjectivity is introduced into the corresponding state vectors and the
Hilbert space through the specification of the inner product. It was necessary
to make an assumption -- a judgement -- about the a priori weight assigned to
any region of configuration space.

It is possible to develop the CAQT approach further and use an entropic
argument \cite{Caticha98c} to show that time evolution is not merely linear
but that it is also unitary. This allows us to introduce observables other
than position. It is also possible to consider more complex many-particle
sources and detectors and thus generalize to many-particle systems. But these
developments lie beyond the limited goal of this paper which was to review the
argument from the objective information embodied in consistent amplitudes to
the Bayesian probabilities of quantum theory.

\end{document}